\documentclass[%
 reprint,
 amsmath,amssymb,showkeys
pra,
]{revtex4-2}

\usepackage{graphicx}
\usepackage{dcolumn}
\usepackage{bm}

\usepackage{blindtext}

\usepackage{caption}
\usepackage{subcaption}
\usepackage{float}
\usepackage{wrapfig}

\begin{document}


\title{Interaction effects on the itinerant ferromagnetism phase transition}
\author{Jordi Pera}
\email{jordi.pera@upc.edu}
\author{Joaquim Casulleras}
\email{joaquim.casulleras@upc.edu}
\author{Jordi Boronat}
\email{jordi.boronat@upc.edu}
\affiliation{%
 Departament de F\'\i sica, Campus Nord B4-B5, Universitat Polit\`ecnica de 
Catalunya, E-08034 Barcelona, Spain
}%

\date{\today}

\begin{abstract}
Itinerant ferromagnetism is one of the most studied quantum phase transitions, 
the transition point and the nature of this phase transition being 
widely discussed. In dilute Fermi liquids, this analysis has been carried out 
up to second-order in the gas parameter, where the results for any spin 
degeneracy are universal in terms of only the $s$-wave scattering length $a_0$. 
We extend this analysis to third-order where energies depend, not only on 
$a_0$, but also on the $s$-wave effective range $r_0$ and the $p$-wave 
scattering length $a_1$. The introduction in the theory of these new parameters 
changes the transition point, with respect to the second-order estimation, and 
also can modify the nature of the phase transition itself. We analyze these 
interaction effects on the phase transition for different spin values. The 
emerging phase diagram shows that the type of ferromagnetic transition changes 
dramatically as a function of $r_0$ and $a_1$ and, importantly, that this 
classification is not solely determined by the spin value, as happens at second 
order. 
\end{abstract}

\maketitle


\section{Introduction}
The Stoner model of itinerant ferromagnetism~\cite{stoner} predicts that a 
dilute paramagnetic Fermi gas with spin $S=1/2$ becomes ferromagnetic by 
increasing the density. At very low density the gas is paramagnetic due to its 
lower Fermi energy. However, as the density increases, the energy of the non-polarized phase surpasses that of the polarized one, due to the increased interaction effects among particles of different spin. In the polarized phase, interparticle interactions are negligible due to the inhibition of $s$-wave collisions. The experimental 
observation of itinerant ferromagnetism in Fermi liquids or gases has been 
extremely difficult. Ultracold Fermi gases have emerged as one of the best 
platforms to reach the conditions for this ferromagnetic transition be 
observed~\cite{pfleiderer,leduc}. However,  at large densities the 
ferromagnetic phase is metastable and competes with the formation of 
dimers~\cite{ji}. In fact, there was a first observation of the 
ferromagnetic transition in cold Fermi 
gases~\cite{jo}, but the work was revised and finally the authors concluded 
that the formation of dimers prevented its observation \cite{sanner}. 
More recently, it was reported that the ferromagnetic 
state was effectively observed in a Fermi $^7$Li gas around a gas parameter 
$x=k_Fa_0\simeq 1$, with $k_F$ the Fermi momentum and $a_0$ the $s$-wave
scattering length~\cite{valtolina}. On the other hand, this transition has
been extensively studied from a theoretical point of view, mainly using quantum 
Monte Carlo 
methods~\cite{conduit,pilati,chang,conduit2,massignan,cui,arias,pilati2}, which 
agree to localize the ferromagnetic transition around $x\simeq1$.

Perturbation theory is a well established theoretical tool to study itinerant 
ferromagnetism. At first 
order, corresponding to the Hartree-Fock approximation, the celebrated Stoner 
model~\cite{stoner} predicts a continuous phase 
transition for $S = 1/2$ and a first-order one for $S > 1/2$. 
Following perturbation theory, at second order 
the Fermi gas suffers a discontinuous transition for any spin value~\cite{Pera}. 
However, second-order perturbation theory is still not accurate enough as it 
cannot reproduce Monte Carlo results close to the predicted critical density. 
In a previous work, the energy as a function of the gas parameter and the spin 
polarization at third-order has been derived \cite{Pera2}. Unlike  second 
order, a fully analytical expression was not possible to obtain. It is worth 
noticing that, going beyond second order, universality breaks, meaning that the 
energy
dependence is no longer solely determined by the $s$-wave
scattering length $a_0$. Two additional scattering parameters come into play in 
the description: the $s$-wave effective range $r_0$ and the $p$-wave scattering 
length $a_1$.

In recent years, SU(N) Fermi gases have been produced experimentally.
For example, 
Ytterbium~\cite{pagano}, with spin 5/2, and Strontium~\cite{goban}, with spin
9/2, are now available for studying Fermi gases with large spin 
degeneracy. This relevant achievement has renewed the theoretical interest on 
these gases.  Cazalilla \textit{et 
al.}~\cite{cazalilla_1,cazalilla_2}  showed  that Fermi gases made of alkaline 
atoms, with two electrons in the external shell, such as $^{173}$Yb, present an 
SU(N) emergent symmetry. Due to this mathematical property, they argued that 
the ferromagnetic transition should be a discontinuous one for $S > 1/2$.
In Ref.~\cite{effect_vary_N}, the same authors studied the interaction effects 
in SU(N) Fermi gases as a function of the spin degeneracy $N$. Collective 
excitations in SU(N) Fermi gases with tunable spin were also investigated in 
Ref.~\cite{collective_excitations}. On the other hand,
the study of the prethermalization of these systems showed that, under some 
conditions, the imbalanced initial state could be 
stabilized for a certain time~\cite{prethermalization}. Recently, the 
thermodynamics of  $^{87}$Sr,  with $N$ that can be tuned up to 10, was 
analyzed in Ref.~\cite{estronci_N_10}. The   
temperature dependence of itinerant ferromagnetism in SU(N) Fermi 
gases, at  second order of perturbation theory, has been studied for the 
first time by Huang and Cazalilla~\cite{cazalilla-t}. 
Once in the fully polarized phase only $p$-wave scattering is possible.
Maki and Enss~\cite{transport_p_wave} have studied the shear viscosity and the 
thermal conductivity of a polarized Fermi gas. More 
recently, Bertaina \textit{et 
al.}~\cite{bertaina_p_wave} reported quantum Monte Carlo calculations 
of single-component ultracold Fermi gases with 
$p$-wave interactions. Finally, and still with single-component Fermi gases, Gao 
\textit{et al.}~\cite{temp_contact_p_wave} established a relationship 
between the finite-temperature loss rate and the $p$-wave contact.

In the present work, we explore finite-range effects, defined in terms of 
the  scattering parameters  $r_0$ and $a_1$,  on the magnetic behavior of 
a Fermi gas. Our goal is to characterize the ferromagnetic transition for a 
range of spin values, from  $S=1/2$ to $S=9/2$, as a function of the 
above-mentioned scattering parameters. Our results show that the nature of the 
phase transition is not uniquely defined by the spin value since, depending 
on the scattering parameters, we can have a continuous transition, a 
discontinuous one or even no transition at all. This richness contrasts with the 
prediction made at second-order perturbation theory, where a unique kind of 
phase transition (first-order phase transition) was found.

\section{Perturbative series}
We study a repulsive Fermi gas at zero temperature, with spin $S$ and spin  
degeneracy $\nu=2 S +1$, using perturbation theory. The energy of the gas is  
a combination of analytic results and numerical estimations.  In the very 
dilute 
gas regime, only particles with different $z$-spin component interact via a 
central potential $V(r)$ ($s$-wave scattering). The first and second order 
terms~\cite{Pera} are universal since the energy depends only on the $s$-wave 
scattering length. However, going to third order in the gas parameter, we 
include 
interactions also between particles of the same $z$-spin component. The 
resulting energy involves not only the $s$-wave scattering length
but also the $s$-wave effective range and the $p$-wave scattering
length~\cite{Pera2},
\begin{widetext}
\begin{equation}
\hspace*{-0.75cm}
\label{eq.forfinal}
\begin{aligned}
     \frac{E}{N}=\frac{3}{5}\epsilon_F\bigg\{ 
\frac{1}{\nu}\sum_{\lambda}C_{\lambda}^{5/3}+\frac{1}{\nu}\sum_{\lambda}\frac{2}
{3\pi}C_{\lambda}^{8/3}(k_Fa_1)^3
   +\frac{5}{3\nu}\sum_{\lambda_1,\lambda_2} \bigg[ \bigg(  
\frac{2}{3\pi}(k_Fa_0)C_{\lambda_1}C_{\lambda_2}+\frac{4}{35\pi^2}C_{\lambda_1}
C_{\lambda_2}\frac{C_{\lambda_1}^{1/3} +C_
{\lambda_2}^{1/3}}{2} \, F(y)(k_Fa_0)^2\\    
+\frac{1}{10\pi}C_{\lambda_1}C_{\lambda_2}\bigg(\frac{C_{\lambda_1}^{2/3}+C_{
\lambda_2}^{2/3}}{2}\bigg)\bigg[\frac{r_0}{a_0}+2\frac{a_1^3}{a_0^3}\bigg]
(k_Fa_0)^3
   +\frac{3}{32\pi^7}\big[E_3+E_4+\sum_{\lambda_3}( 
(2-3\delta_{\lambda_1,\lambda_3} 
-3\delta_{\lambda_2,\lambda_3})E_5) \big](k_Fa_0)^3\bigg)(1-\delta_{\lambda_1, 
\lambda_2})\bigg] \bigg\} \ .
\end{aligned}
\end{equation}
\end{widetext}
The number of particles in each spin channel is 
$N_\lambda=C_\lambda N/\nu$, with $N$ the total number of particles and   
$C_\lambda$ being the fraction of $\lambda$ particles (normalized to be one if 
the 
system is unpolarized, $N_\lambda=N/\nu$, $\forall \, \lambda$).  The Fermi  
momentum  is $k_{F,\lambda}=k_F C_\lambda^{1/3}$, $k_F$ being 
$(6\pi^2n/\nu)^{1/3}$, and $\epsilon_F=\hbar^2 k_F^2/ 2m$ being the Fermi 
energy. In Eq. (\ref{eq.forfinal}), the terms $E_3$, $E_4$, and $E_5$ cannot 
be derived analytically and requires of a numerical 
integration~\cite{Pera2,Bishop}.  

Our goal is the description of the ferromagnetic phase transition as a function 
of the spin and the scattering parameters entering in Eq. (\ref{eq.forfinal}). 
The phase diagrams are  obtained by minimizing the energy for any value of 
$r_0$ and $a_1$ 
as a function of the gas parameter $x=k_Fa_0$. After this minimization, one can 
get the polarization $P$ in terms of $x$. By analyzing this dependence one 
can study the nature of the transition for each set of scattering parameters. 
For $S>1/2$ there is not a single definition of the spin polarization. In our 
study, we assume that, if the  concentration of one species increases, the 
concentration of all the other species diminish equally.


\section{Results}
\begin{figure*}
    \centering
    \includegraphics[width=1\linewidth]{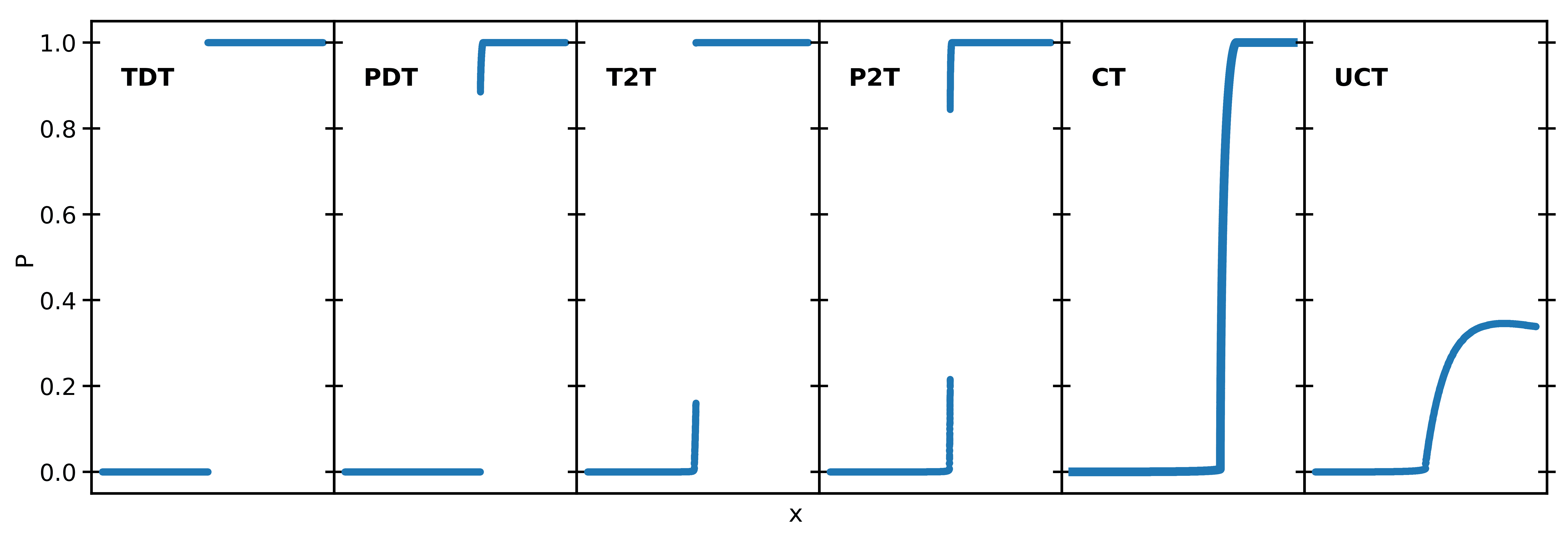}
    \caption{Main types of phase transitions using the polarization $P$ as 
order parameter. From left to right, discontinuous, double and 
continuous.}
    \label{fig:pols}
\end{figure*}

We explore the nature of the phase transition for different values of the 
$p$-wave 
scattering length and $s$-wave effective range. Given certain values for $a_1$ 
and $r_0$, we analyze how the polarization corresponding to the minimum 
energy changes in terms of the gas 
parameter.  Our results shows a rich variety of different ferromagnetic 
transitions and even situations where this transition disappears.
In Fig.~\ref{fig:pols}, we illustrate these different scenarios. TDT and CT 
correspond to a total discontinuous (first-order) phase transition and to a 
continuous (second-order) phase transition, respectively. PDT stands for discontinuous transitions with gaps in the polarization smaller than 
one. Surprisingly, there can be cases which experience a kind of double transition. First, the polarization increases continuously from 0 to a certain value smaller than 1. And then, a finite jump is observed as if it was a discontinuous transition. These situations are labelled as T2T and P2T. In T2T, after the finite jump, the polarization reaches 1; while in P2T, the polarization is still smaller than 1. The case UCT corresponds to a continuous transition but to states 
that are nut fully polarized, meaning that in these scenarios no 
itinerant ferromagnetism is possible. The double transitions and 
the suppression of the ferromagnetic transition  are observed only at 
third-order of perturbation theory. In fact, at second-order the transition is 
first-order for all values of the spin $S$~\cite{Pera}. 

\begin{figure}[h]
    \centering
    \includegraphics[width=1\linewidth]{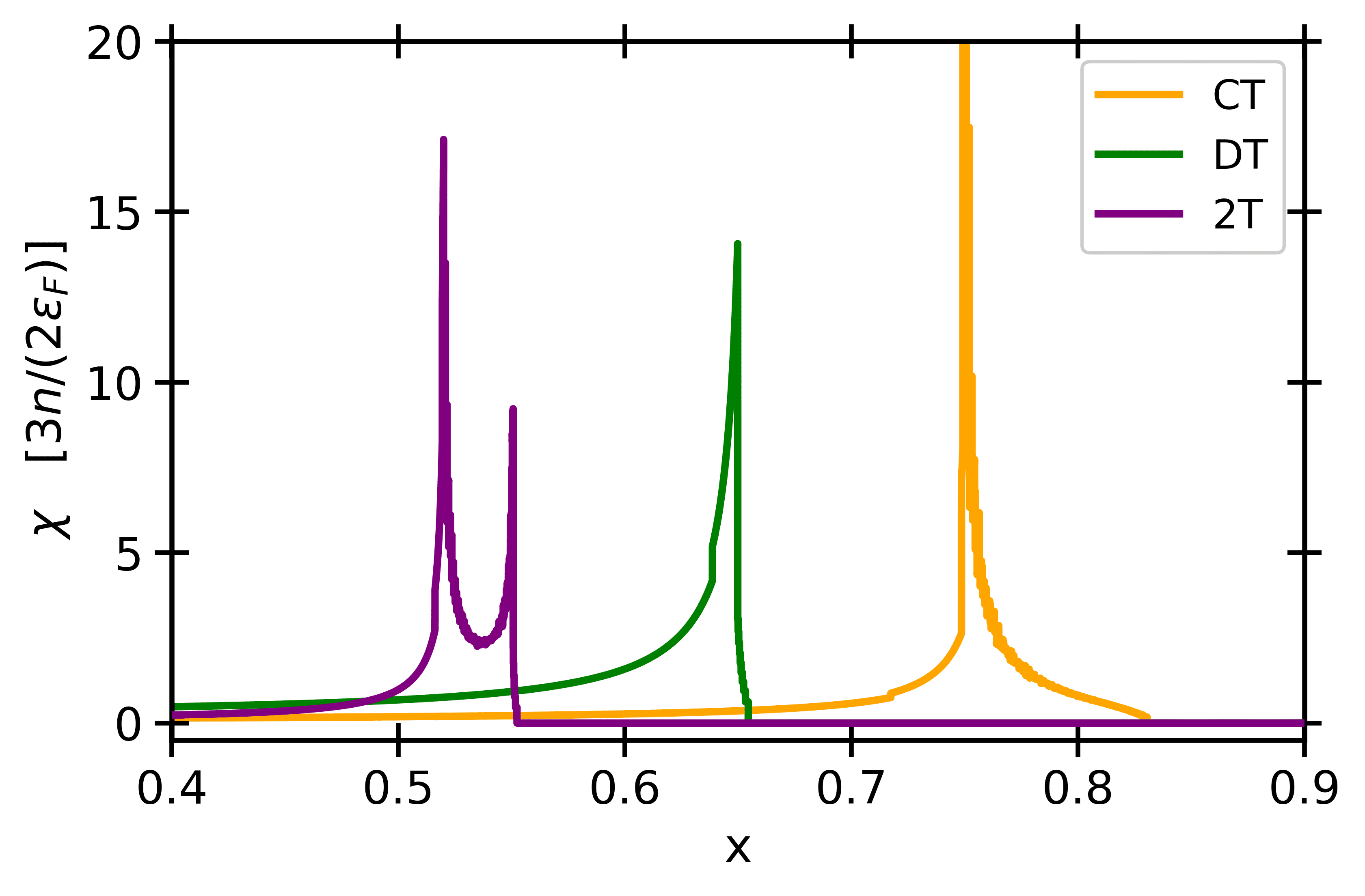}
    \caption{Magnetic susceptibilities at the phase transition. From left to 
right: double, discontinuous, and continuous transitions. Notice 
that the continuous one diverges at the transition point.}
    \label{fig:susc}
\end{figure}

The magnetic susceptibility $\chi$, defined as
\begin{equation}
   \frac{1}{\chi}=\frac{1}{n}\left(\frac{\partial^2 (E/N)}{\partial 
P^2} \right)_x \ ,
\label{suscept}
\end{equation}
can be obtained from the energy (\ref{eq.forfinal}). In Fig.~\ref{fig:susc}, 
we show characteristic results of $\chi$ around the transition point for the 
main types of behaviors: double, discontinuous, and continuous. The location of the peaks indicates the transition points, their 
height being finite and infinite for the discontinuous and continuous case, 
respectively.

\begin{figure}[t]
    \centering
    \includegraphics[width=1\linewidth]{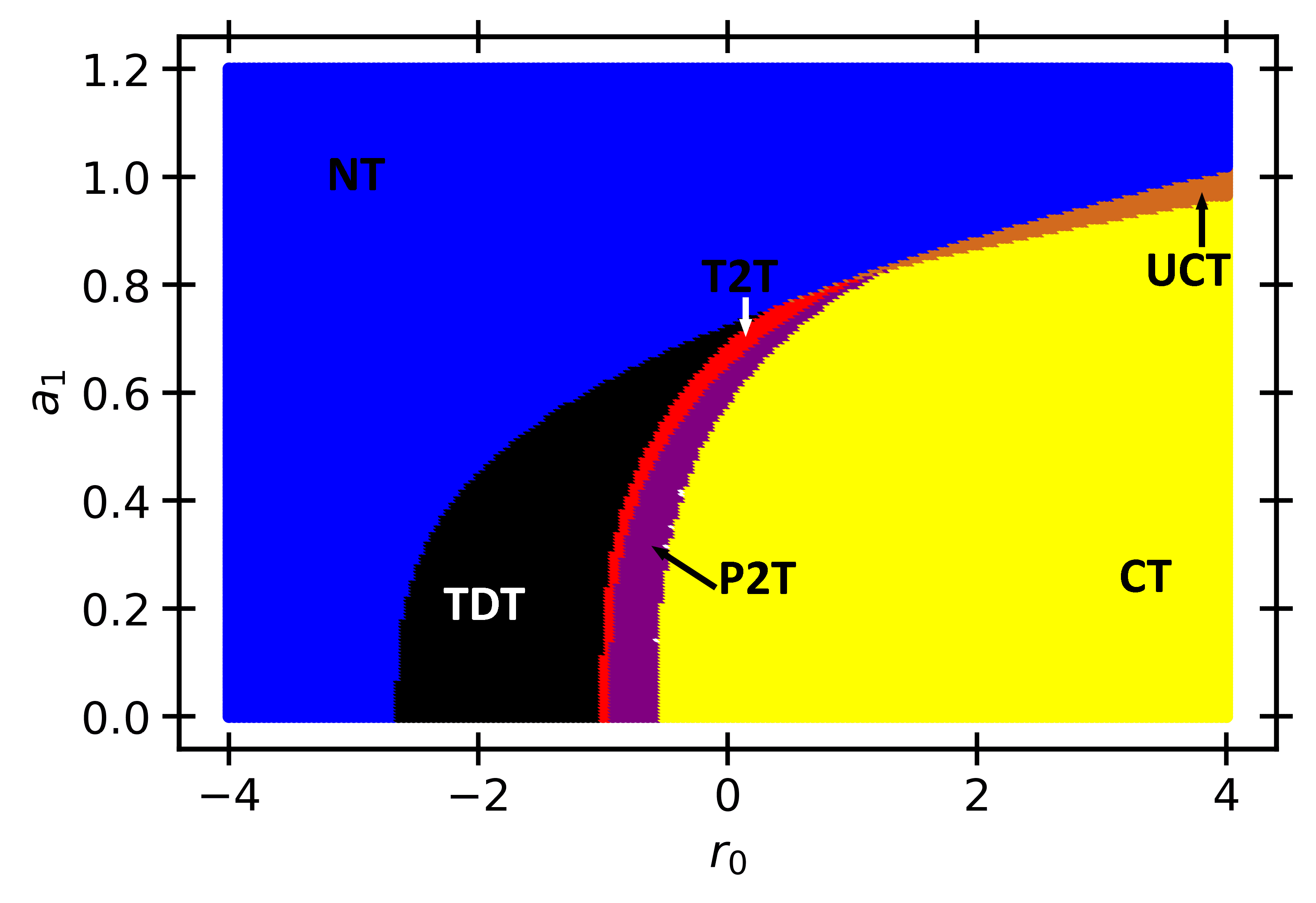}
    \caption{Ferromagnetic phase transition types for $S=1/2$.}
    \label{fig:exv2}
\end{figure}

In Fig.~\ref{fig:exv2}, we show the different types of phase transition for 
$S=1/2$ as a function of $r_0$ and $a_1$. We selected a range 
of values for $r_0$ and $a_1$ where the diagram is outstanding. With different 
colors, and the acronyms discussed above, 
we can easily identify the different regimes. As one can see, 
the dominant transition is a continuous one. However, for small and negative 
$r_0$ values there are discontinuous transitions. And also, if $r_0$ is large 
and 
negative or $a_1$ is large, there is no transition at all (NT), i.e, , the 
Fermi gas remains unpolarized for any value of the gas parameter. We 
point out that, if $p$-wave interactions between particles of same spin are 
included, the hard-spheres case ($r_0=2/3$ and $a_1=a_0$) falls into this 
latter case.


\begin{figure*}[]
    \centering
    \begin{subfigure}[b]{0.45\textwidth}
        \centering
        \includegraphics[width=\textwidth]{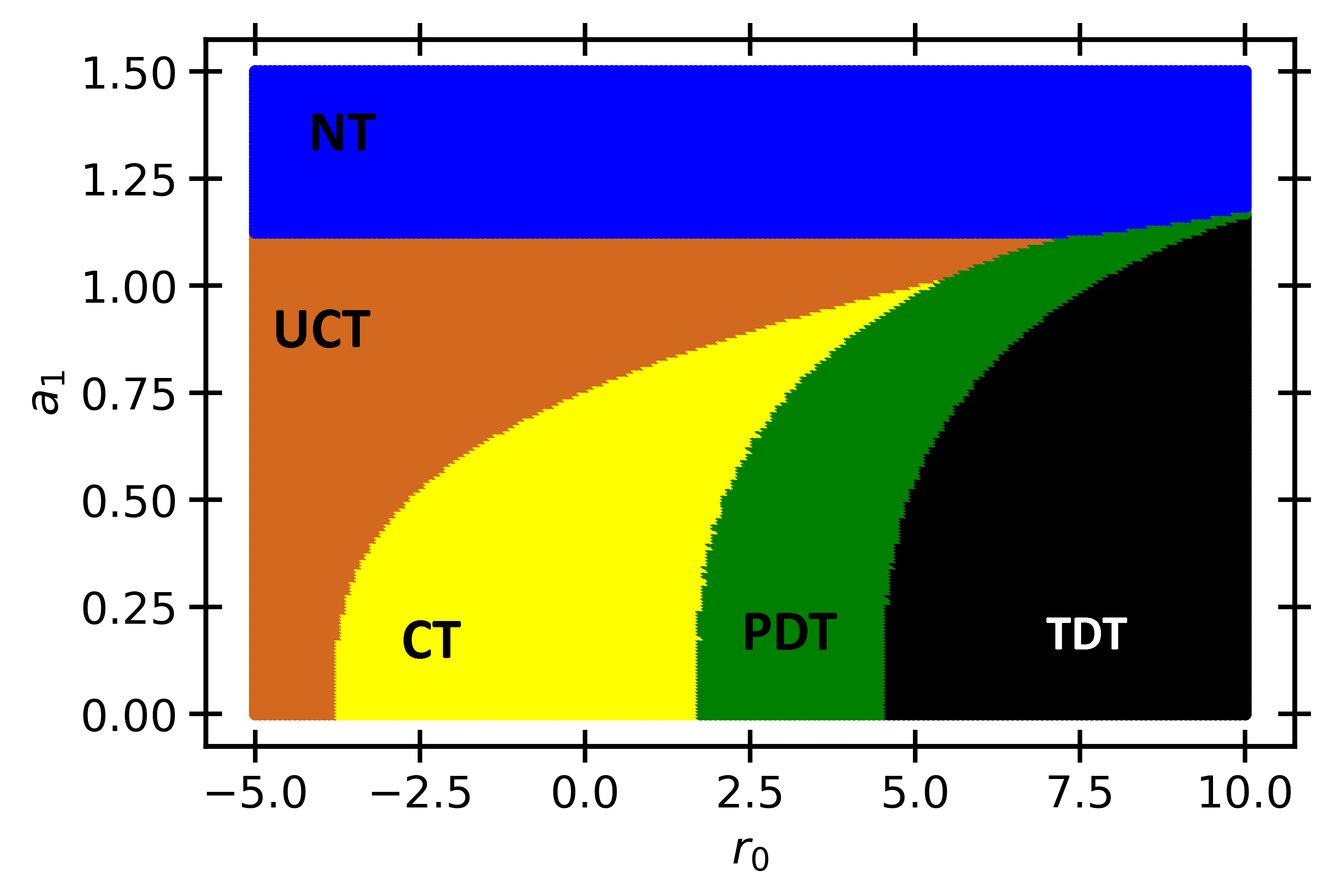}
    \end{subfigure}
    \begin{subfigure}[b]{0.45\textwidth}
        \centering
        \includegraphics[width=\textwidth]{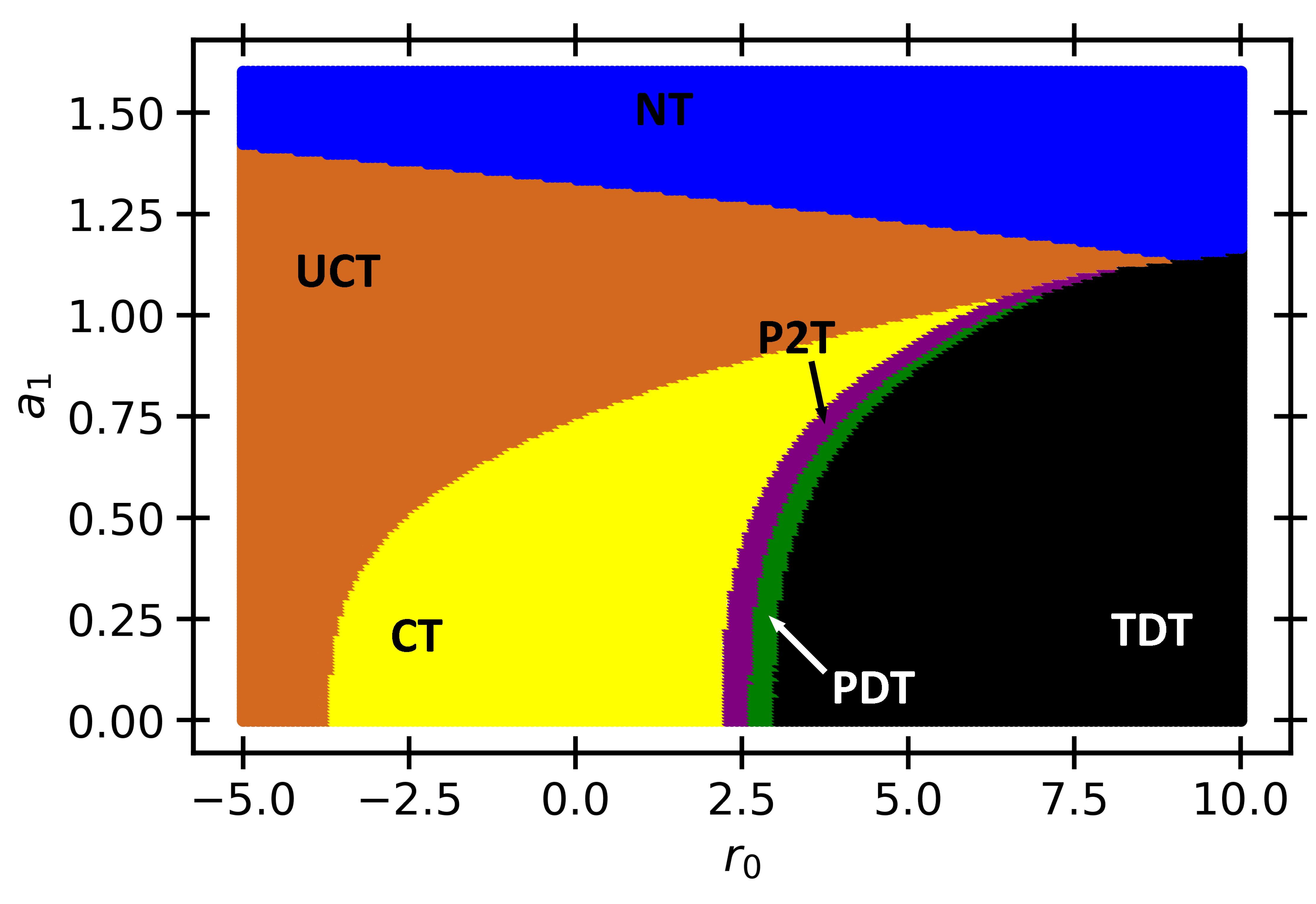}
    \end{subfigure}
    \\
    \begin{subfigure}[b]{0.45\textwidth}
        \centering
        \includegraphics[width=\textwidth]{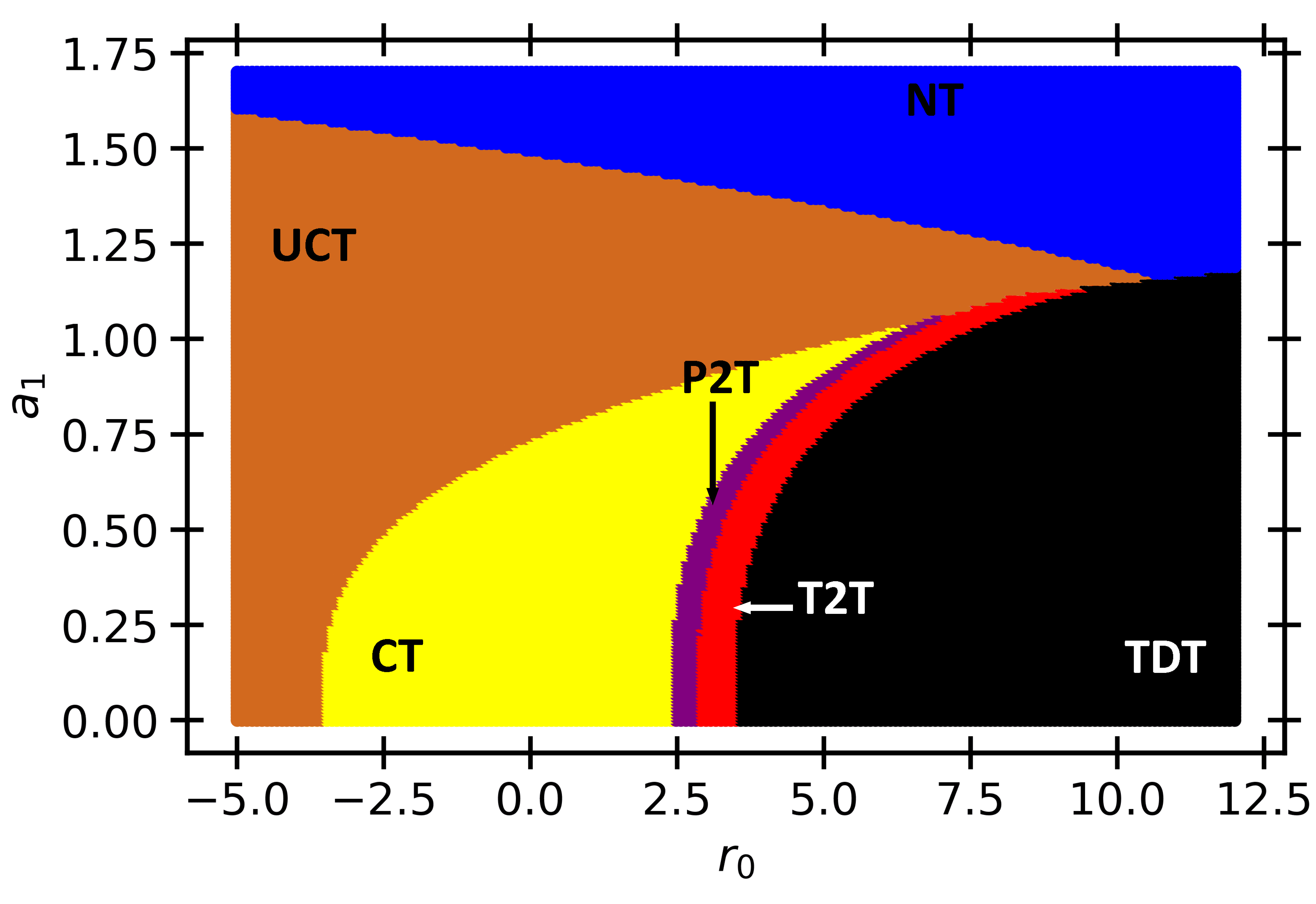}
    \end{subfigure}
    \begin{subfigure}[b]{0.45\textwidth}
        \centering
        \includegraphics[width=\textwidth]{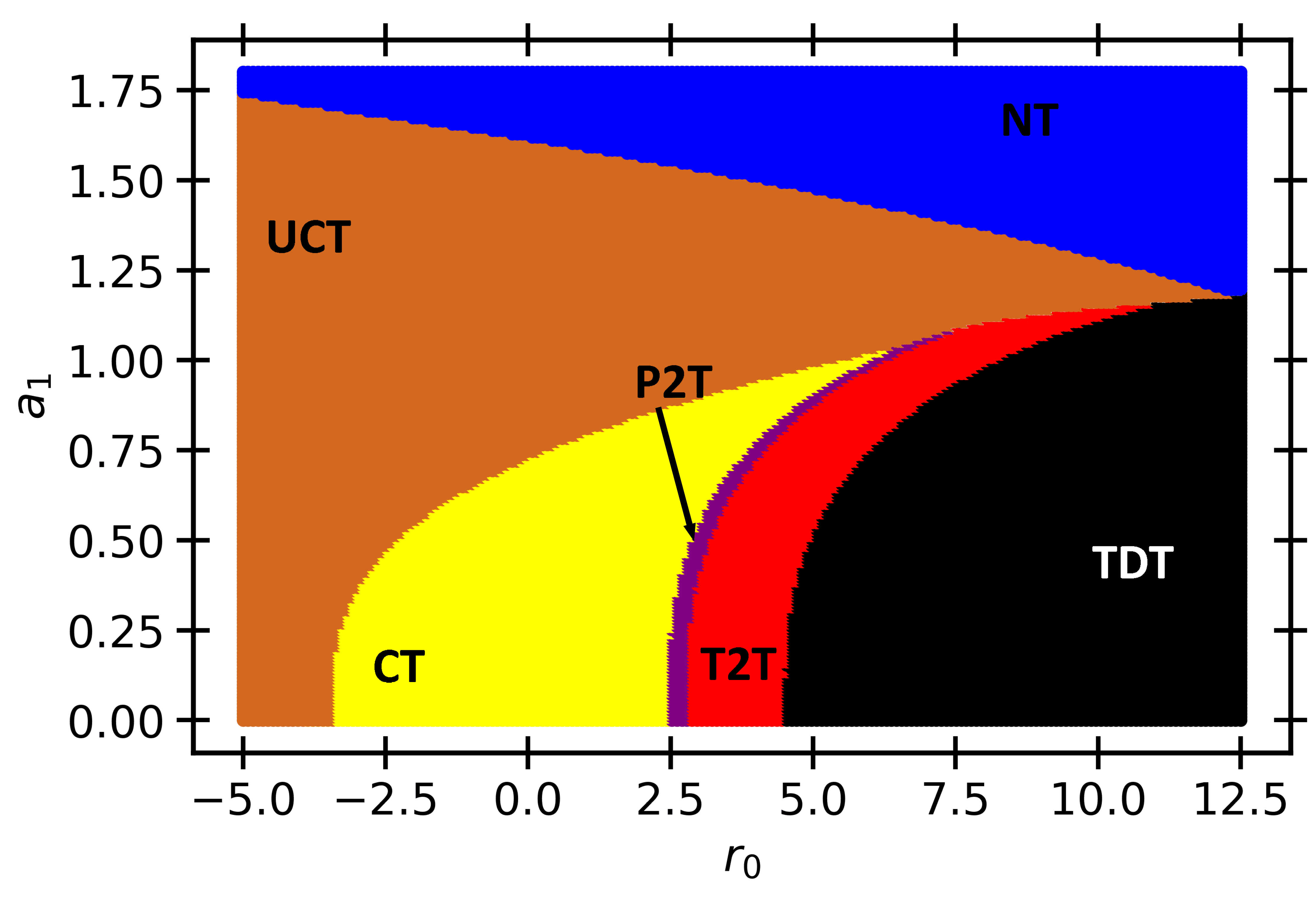}
    \end{subfigure}
    \caption{Ferromagnetic phase transition types for 
$S=3/2$, $5/2$, $7/2$, and $9/2$ (from left to right and from top to bottom).}
\label{fig.explor}
\end{figure*}

Once we increase the spin degeneracy, we continue to observe different 
scenarios. Our results for  $S=3/2$, $5/2$, $7/2$, and $9/2$  are shown in  
Fig.~\ref{fig.explor}. The most relevant difference between $S=1/2$ and $S>1/2$ 
is that the types of phase transitions are distributed oppositely. In spin 
$1/2$, we have continuous transitions for positive $r_0$ and the 
discontinuous ones appear for negative $r_0$ values. Now, for larger spin, the 
situation is reversed. The continuous transitions happen for negatives or small 
$r_0$, and the discontinuous ones for large values of $r_0$. As before, if $a_1$ 
is large enough, there is no transition at all. Therefore, at third order of 
perturbation theory the value of the spin does not determine the phase 
transition type, similarly to the case of $S=1/2$.

There is one specific case which has been studied using diffusion Monte Carlo 
(DMC)~\cite{pilati}  that we want to highlight. In those calculations, a 
hard-sphere potential was considered ($r_0=2a_0/3$ and $a_1=a_0$), but the 
$p$-wave interaction effects between particles of same spin were neglected. If 
we look at Eq. (\ref{eq.forfinal}), the neglected term is the second one in the 
formula, that is, 
\begin{equation}
    \frac{1}{\nu}\sum_{\lambda}\frac{2}{3\pi}C_{\lambda}^{8/3}(k_Fa_1)^3 \ .
\end{equation}
To translate this case into our formalism, one can re-define the effective 
range as $\bar{r}_0/a_0 = (r_0/a_0+2a_1^3/a_0^3)$. In practice, it means to 
follow the line $a_1=0$ with $\bar{r}_0$ in Figs. \ref{fig:exv2} and 
\ref{fig.explor}. For the hard-sphere potential,  
$\bar{r}_0/a_0=2/3+2\approx2.67$.
 Around this region, we have partial 
discontinuous transitions for $S=3/2$ and $5/2$. And, although it is hard to see, for $S=7/2$ and 
$9/2$, there are partial double transitions. This particular case has been 
analyzed in Ref.~\cite{Pera}.

\section{Landau Theory}
Landau theory of phase transitions helps to understand the origin of the 
different transition types observed in our study and the location of the 
critical points. According to the order of our approximation, the parameters 
of the Landau expansion depend on $x$ up to third order. It is written as an 
expansion in terms of the polarization $P$,
\begin{equation}
\begin{aligned}
    f(x,P)-f_0(x)=-\frac{A}{2} (\overline{x}(x)-x_0) P^2-\frac{B(x)}{3}{|P|}^3
    \\+\frac{C(x)}{4}P^4+\frac{L(x)}{4}P^4\ln{|P|} \ ,
\end{aligned}
    \label{Elandauv}
\end{equation}
with $f(x,P)= 5E/(3N\epsilon_F)$ and $x_0=\pi/2$. The coefficients in Eq. 
(\ref{Elandauv}) depend on the gas parameter $x$ through an involved expression 
(see Appendix~\ref{appendix:b}). Although it is not 
shown here, we point out that the parameter $L$ is the only 
\textit{universal} one since it does not depend on $r_0$ and $a_1$. For 
the sake of simplicity, from now on, we will not specify that the coefficients 
of Eq. (\ref{Elandauv}) depend on the gas parameter $x$. In order to determine 
the transition point, one imposes two conditions: \textit{i})  the energy has to 
be a minimum, that is, its first 
derivative must be zero, and  \textit{ii}) the energy must be smaller than the 
unpolarized phase energy, to have a global minimum. One solution to these 
criteria is $P^*=0$ and $\overline{x}(x)=x_0$, which would represent a 
continuous transition. However, there are other solutions.

The other solutions, that give rise to discontinuous transitions, satisfy 
the following equations,
\begin{equation}    
\ln{|P^{\ast}|}=-\bigg(\frac{1}{2}+\frac{C}{L}\bigg)+\frac{2B}{3L|P^{\ast}|}  \ 
,
    \label{pstar2nd}
\end{equation}
\begin{equation}
   \overline{x}^{\ast}=x_0-\frac{B}{3A}|P^{\ast}|-\frac{L}{4A}|P^{\ast}|^2 \ .
\label{xstar2nd}
   \end{equation}
We rewrite Eq.~(\ref{pstar2nd}) to better explore the range where
a solution exists,
\begin{equation}
    \frac{\alpha}{|P^{\ast}|}-\ln{|P^{\ast}|}=\beta \ ,
\end{equation}
with $\alpha=2B/(3L)$ and $\beta=1/2+C/L$.
There are three domains in which we find a solution: a) if 
$\alpha\geq0$,  $\beta\geq\alpha$; b) if $\alpha\in(-1,0)$, 
$\beta\leq-1-\ln{(-\alpha)}$; and c) if $\alpha\leq-1$, 
$\beta\leq\alpha$.

We first analyze $S=1/2$, as it is the most relevant case. For this spin value, 
$B=0$, and hence, $\alpha=0$ too. If there is no logarithmic term ($L=0$) and 
$C>0$ (case that corresponds to the Stoner model~\cite{stoner}), the 
Fermi gas shows a continuous transition. However, if a 
logarithmic term with a positive coefficient ($L>0$) is present, then a discontinuous 
transition occurs; this is what happens when we expand the energy up to second 
order in the gas parameter~\cite{Pera}. In the present case, we need to 
analyze the shape of $L$, which is universal, and the effects 
of $r_0$ and $a_1$ on the other parameters in Eq. (\ref{Elandauv}).

\begin{figure}[h]
    \centering
    \begin{subfigure}[b]{0.238\textwidth}
        \centering
        \includegraphics[width=\textwidth]{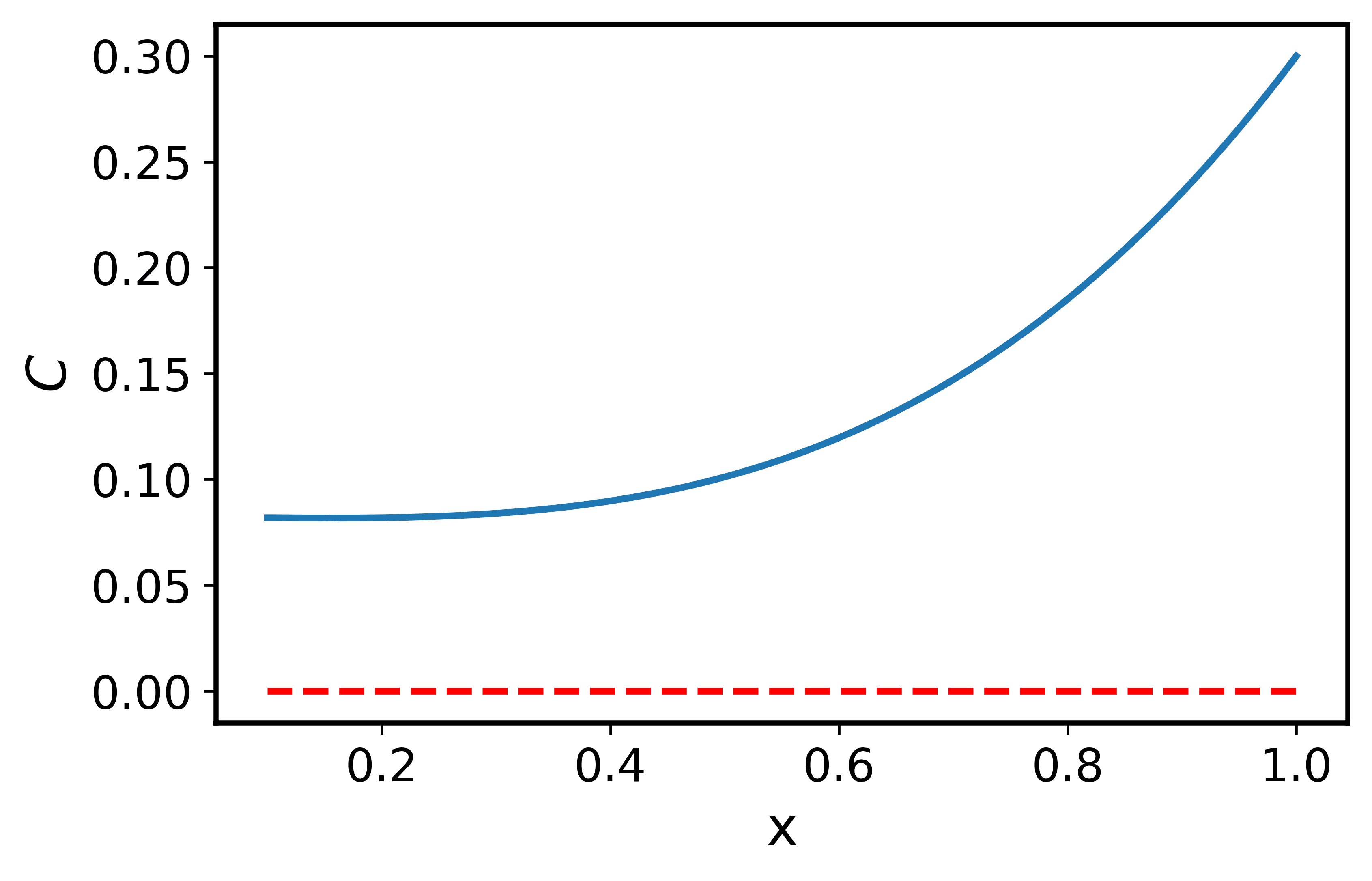}
    \end{subfigure}
    \begin{subfigure}[b]{0.238\textwidth}
        \centering
        \includegraphics[width=\textwidth]{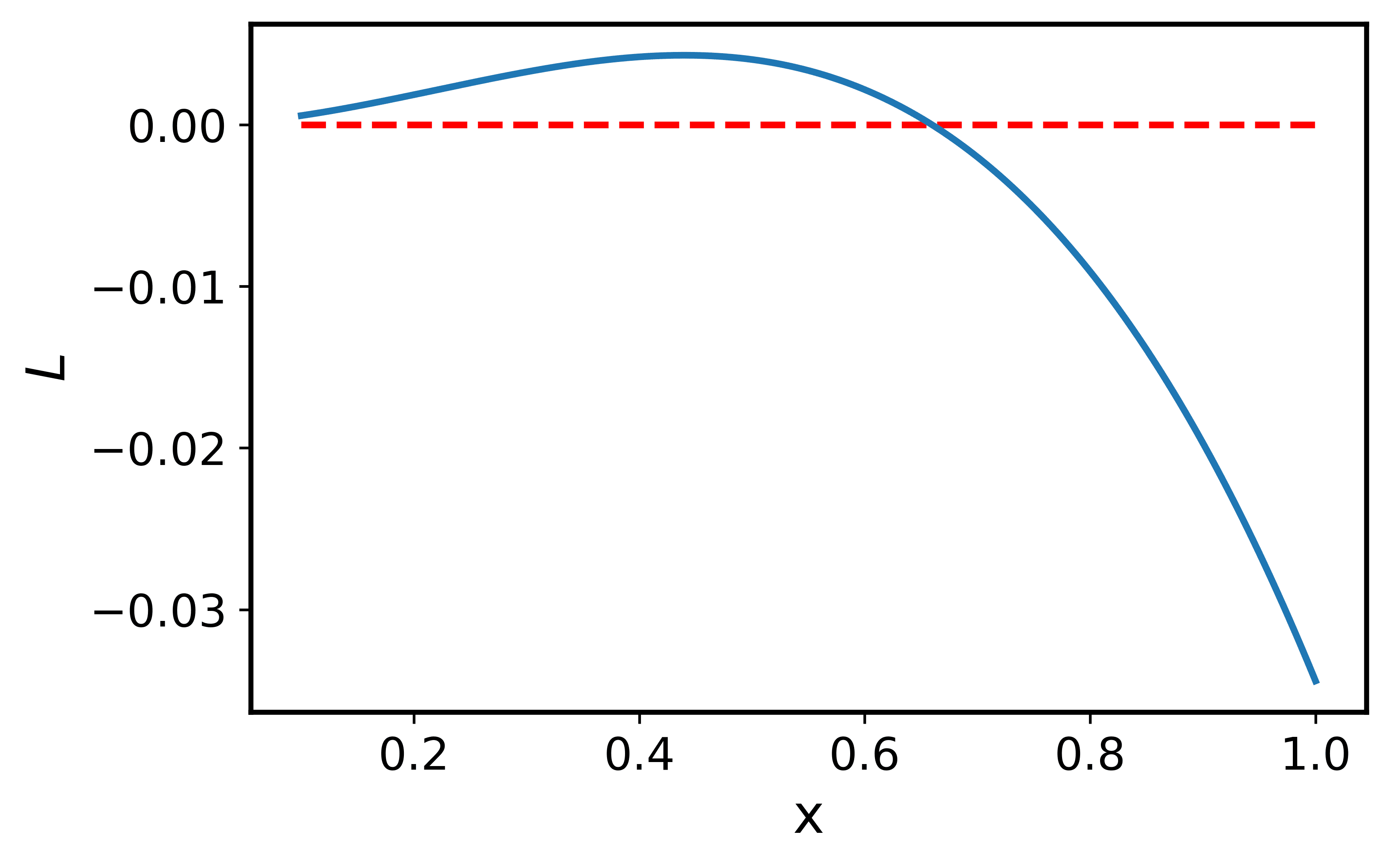}
    \end{subfigure}
    \\
    \begin{subfigure}[b]{0.238\textwidth}
        \centering
        \includegraphics[width=\textwidth]{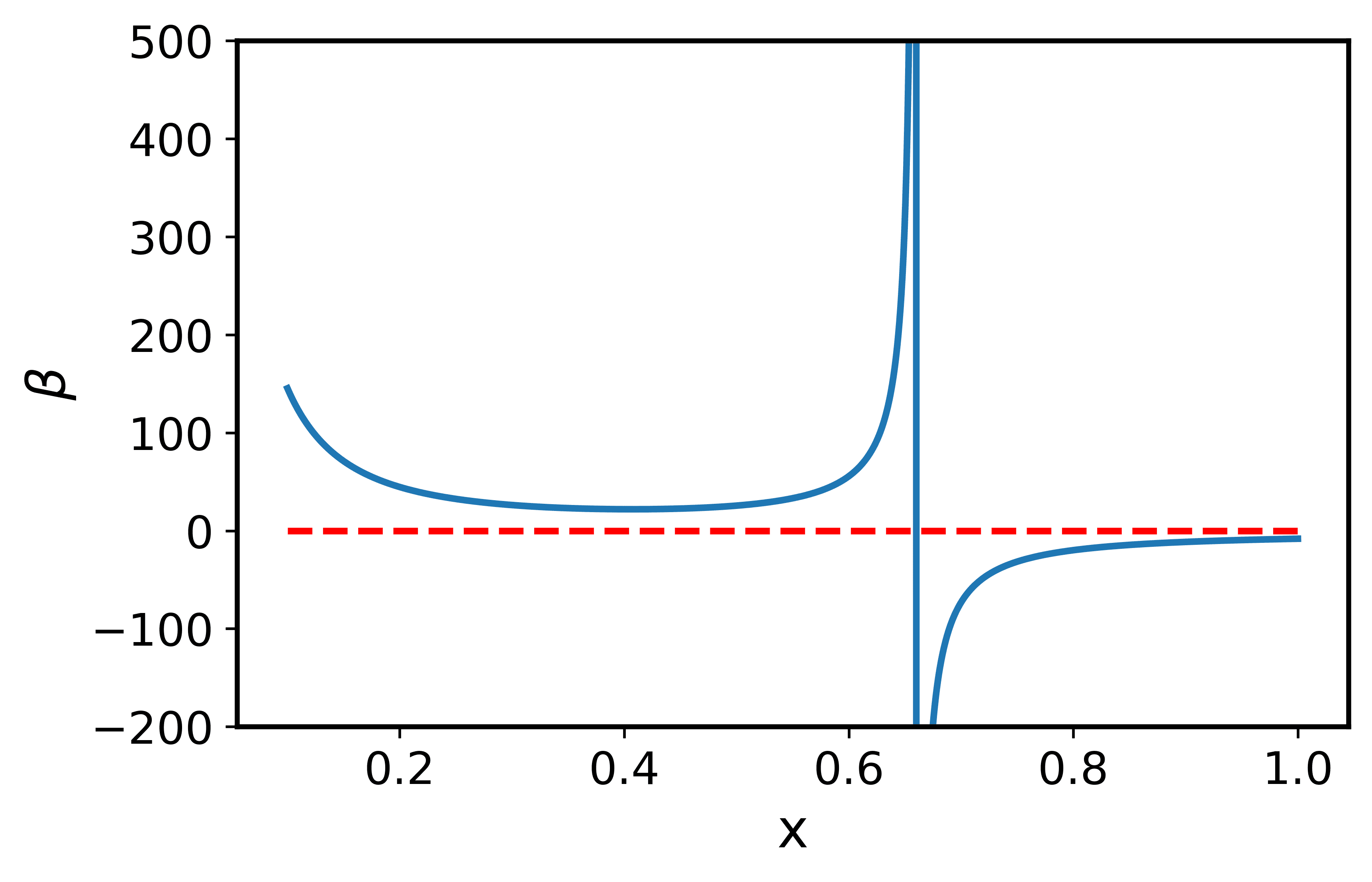}
    \end{subfigure}
    \begin{subfigure}[b]{0.238\textwidth}
        \centering
        \includegraphics[width=\textwidth]{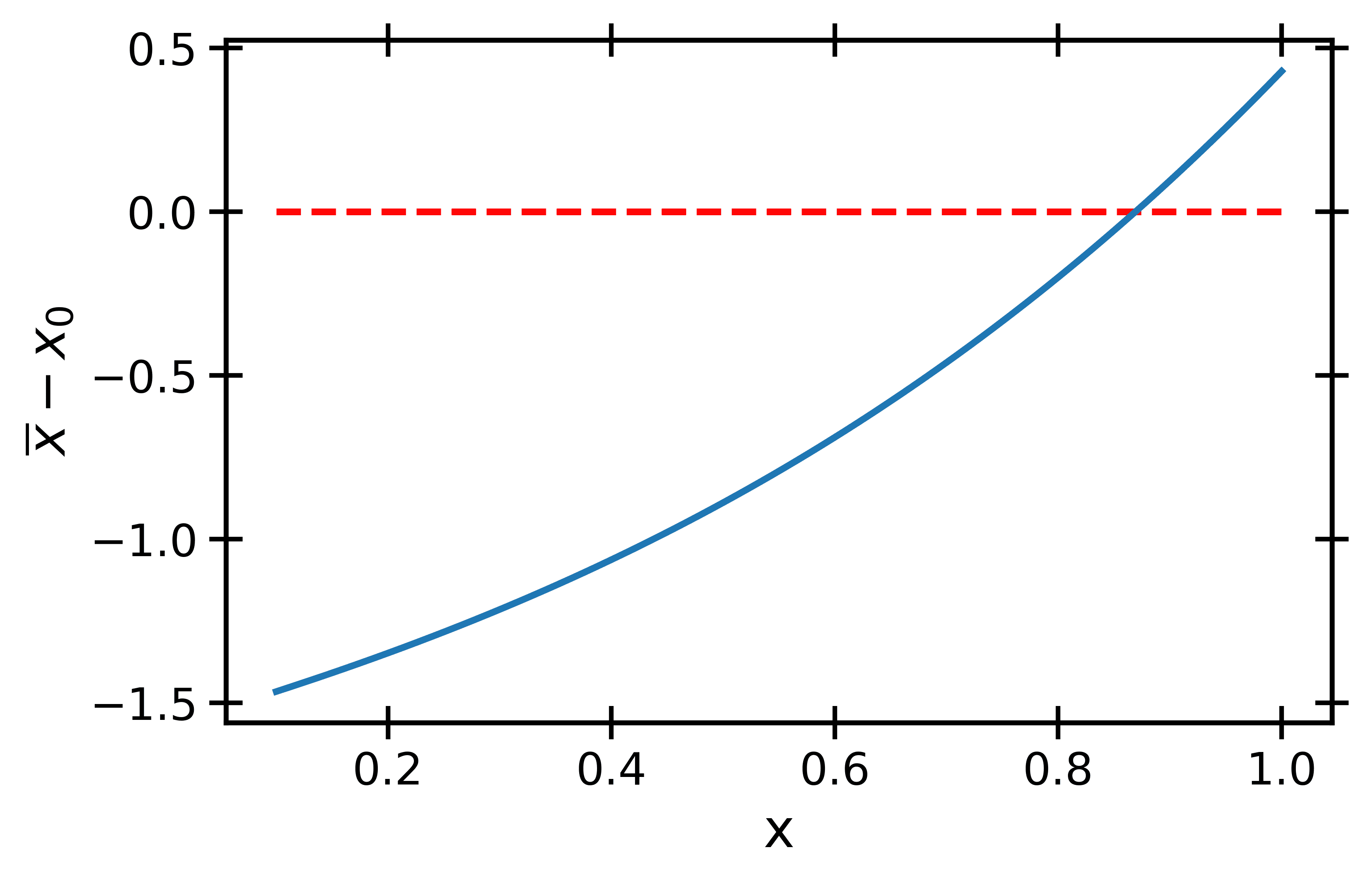}
    \end{subfigure}
    \caption{Parameters $C$, $L$, $\beta$, and 
$\overline{x}(x)-x_0$ in terms of the gas parameter for $S=1/2$, $r_0=2+2/3$ and 
$a_1=0$.}
\label{cas_zero}
\end{figure}

As an example, we plot in Fig. (\ref{cas_zero}) the case exposed at the end of Sec. III, that is, $r_0\approx2.67$ and $a_1=0$. We see that the parameter $C$ is positive whereas 
 $L$ changes its sign. At low values of the gas parameter, $L$ is positive
making a 
discontinuous transition possible. While $x$ grows, it becomes 
negative, implying a continuous transition. In fact, for a discontinuous transition to exist, it is necessary that Eqs. (\ref{pstar2nd}) and (\ref{xstar2nd}) are satisfied before 
$L$ changes sign. It turns out that there is no solution in this regime, hence, 
one finds a critical point $x^*=0.868$ for a continuous transition with
$\overline{x}(x)=x_0$. Summarizing, the transition is continuous, because $L$  has become negative, eliminating the point of inflection in the energy that could cause a discontinuous one.

\begin{figure}[h]
    \centering
    \includegraphics[width=0.9\linewidth]{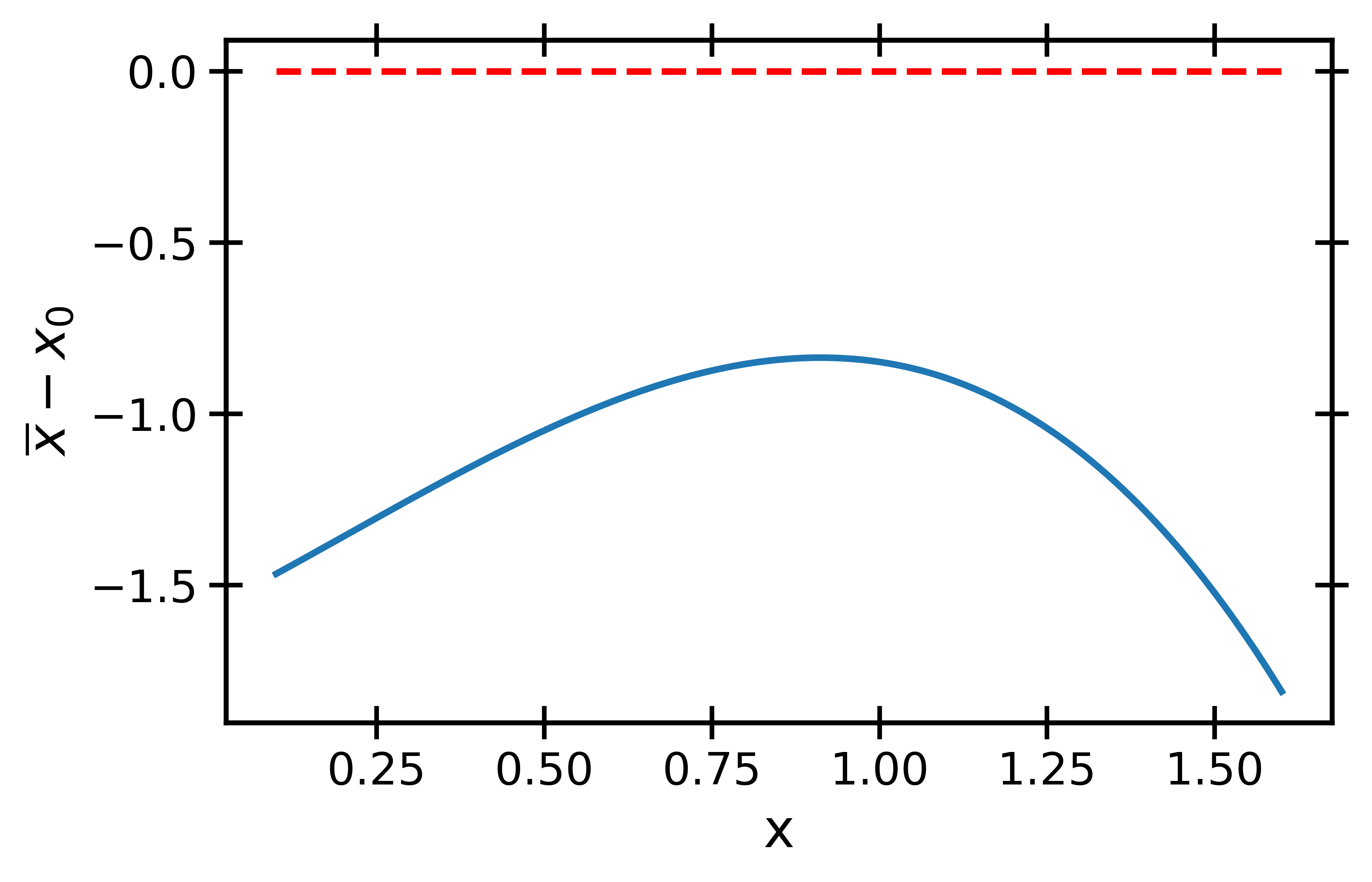}
    \caption{$\overline{x}(x)-x_0$ in terms of the gas parameter for $S=1/2$, $r_0=-5$ and $a_1=0$.}
    \label{cas_u}
\end{figure}

Landau theory also helps to understand the regime where a phase 
transition does not occur, that is, the region depicted in blue in 
Figs.~\ref{fig:exv2} and 
\ref{fig.explor}. In this regime, the shape of the parameters is very similar to 
the 
ones presented in Fig.~\ref{cas_zero}, except for $\overline{x}(x)-x_0$. 
In Fig.~\ref{cas_u}, we show $\overline{x}(x)-x_0$ for $r_0=-5$ and $a_1=0$, a set of values for which there is no transition. 
We can see that that the function $\overline{x}(x)-x_0$ does not cross zero, and hence, there is 
no solution that can give us a transition.

This approach shows also some limitations. Coming back to 
Fig.~\ref{fig:exv2}, one can see that there is a region featuring discontinuous 
transitions when $r_0$ is negative. In principle, the Landau model could be applicable in such cases, because, with negative $r_0$, $C$ 
may become also negative, and then we could have a positive value for $\beta$ 
after 
$L$ changes sign, hence predicting a discontinuous transition. 
However, $C$ becomes negative for a very large value of $r_0$ which is far 
from the results shown in Fig.~\ref{fig:exv2}. Therefore, the Landau expansion 
is only accurate when we deal with 
continuous transitions. This is an expected result since  
the Taylor expansion underlying the theory works effectively for $P \lesssim 
0.5$.

\begin{figure}[h]
    \centering
    \includegraphics[width=0.9\linewidth]{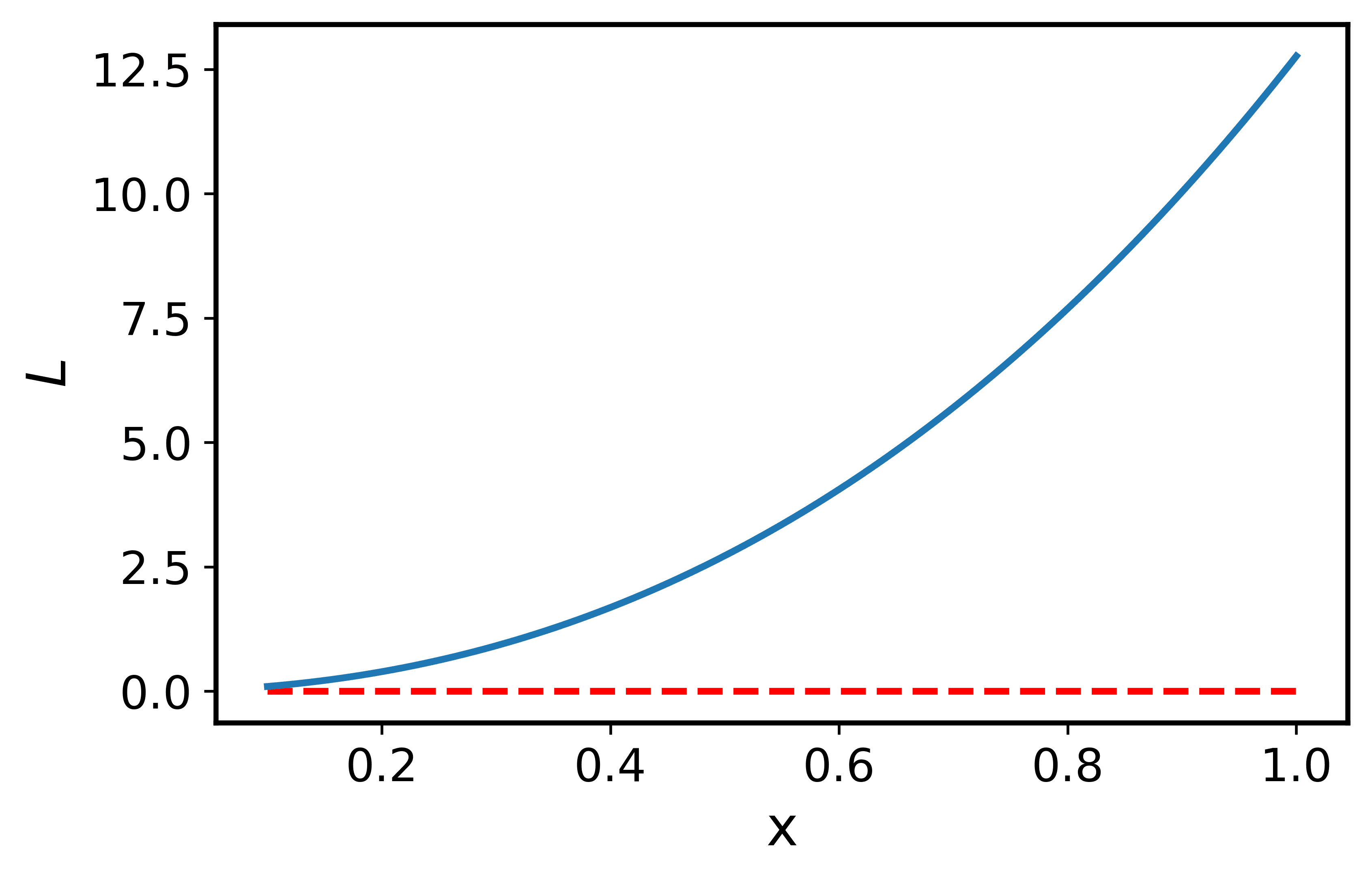}
    \caption{$L$ in terms of the gas parameter for $S=5/2$.}
    \label{cas_L}
\end{figure}

In the following, we analyze the case of $S>1/2$. In this case, the key point 
is to understand the regime of a continuous transition since this is unexpected, as there is a cubic term in the Landau expansion which would lead to a discontinuous transition. If instead one observes a continuous scenario, it must be 
the result of a cancellation between the $P^3$ and 
$P^4 \ln{|P|}$ terms. In Fig.~\ref{cas_L}, we show $L$, which is 
universal. For $S=1/2$, the sign of $L$ is the key to understand what is 
happening as it caused the change of sign of $\beta$. Now, for large $S$, we 
see 
that it is positive everywhere, hence, the origin of the different transitions 
must come either from $B$ or $C$.

\begin{figure}[h]
    \centering
    \begin{subfigure}[b]{0.238\textwidth}
        \centering
        \includegraphics[width=\textwidth]{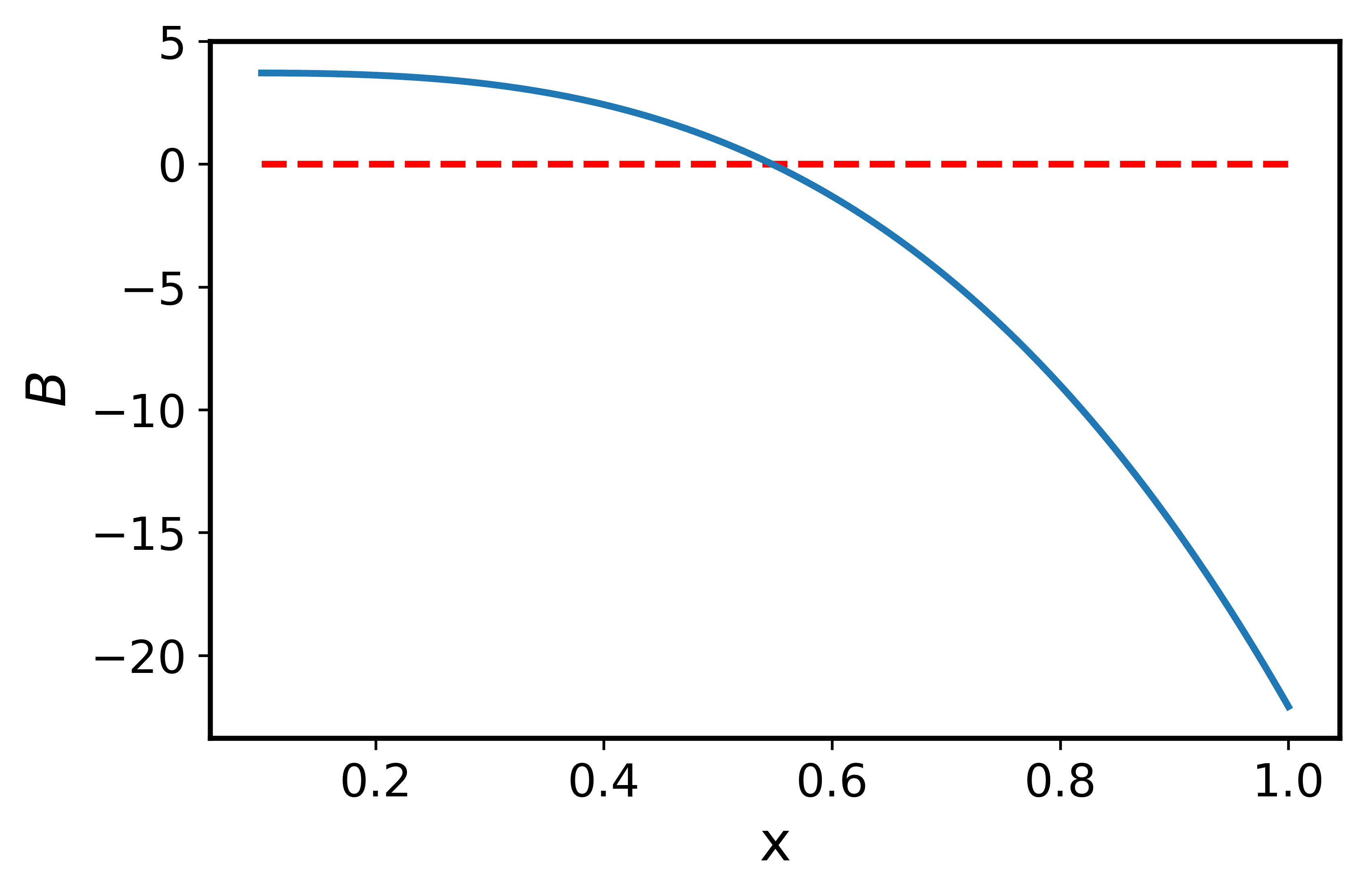}
    \end{subfigure}
    \begin{subfigure}[b]{0.238\textwidth}
        \centering
        \includegraphics[width=\textwidth]{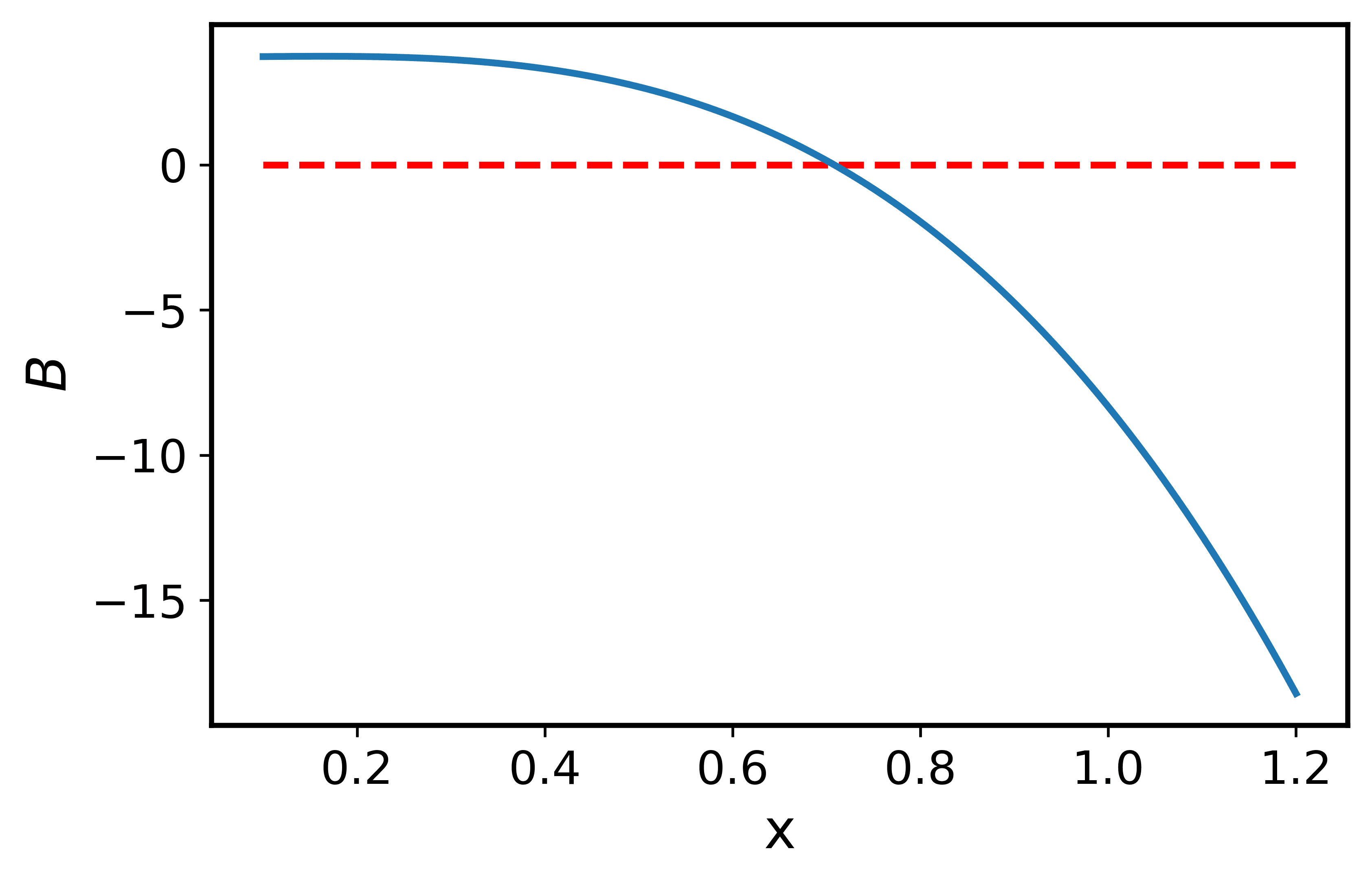}
    \end{subfigure}
    \\
    \begin{subfigure}[b]{0.238\textwidth}
        \centering
        \includegraphics[width=\textwidth]{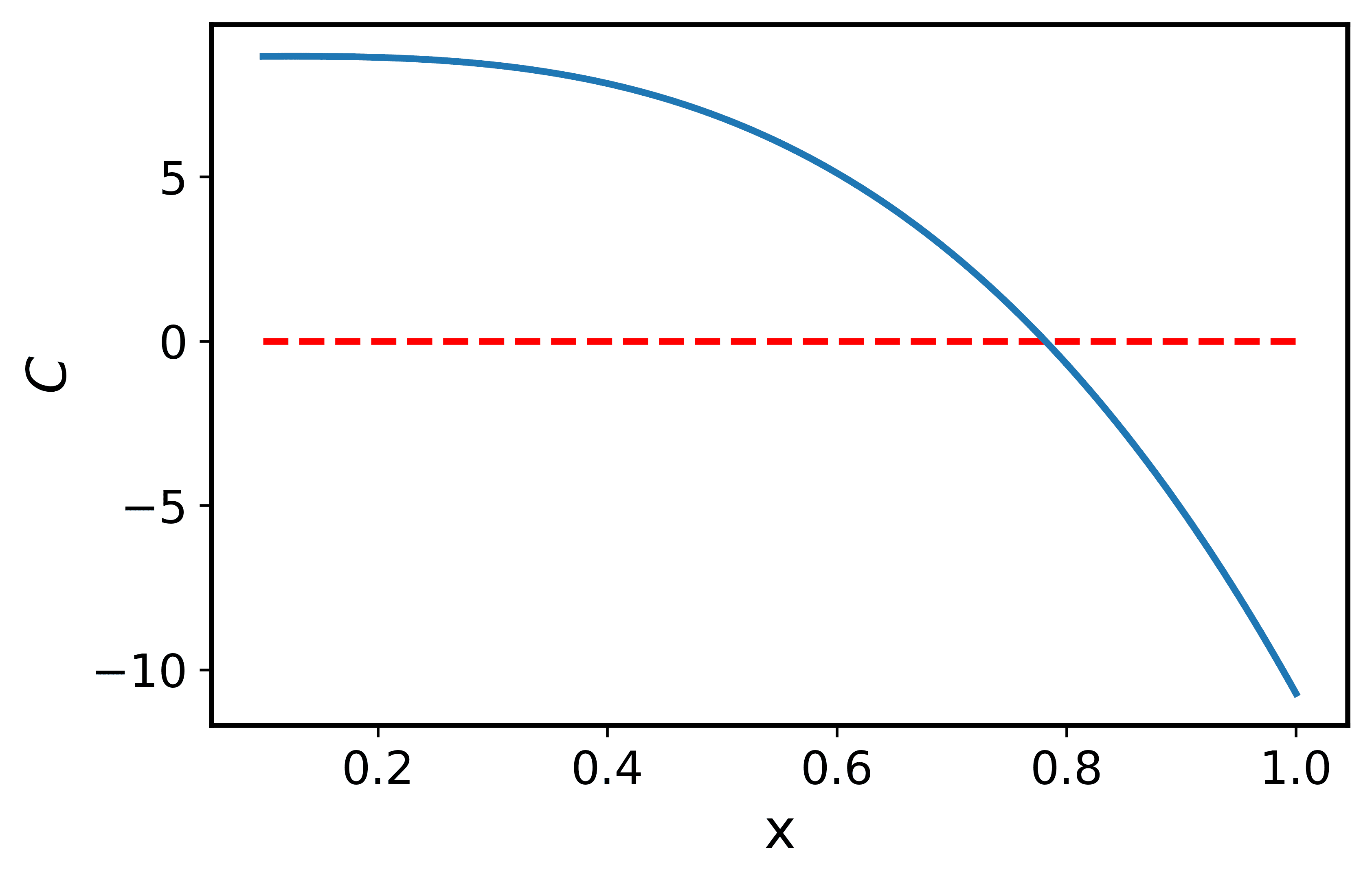}
    \end{subfigure}
    \begin{subfigure}[b]{0.238\textwidth}
        \centering
        \includegraphics[width=\textwidth]{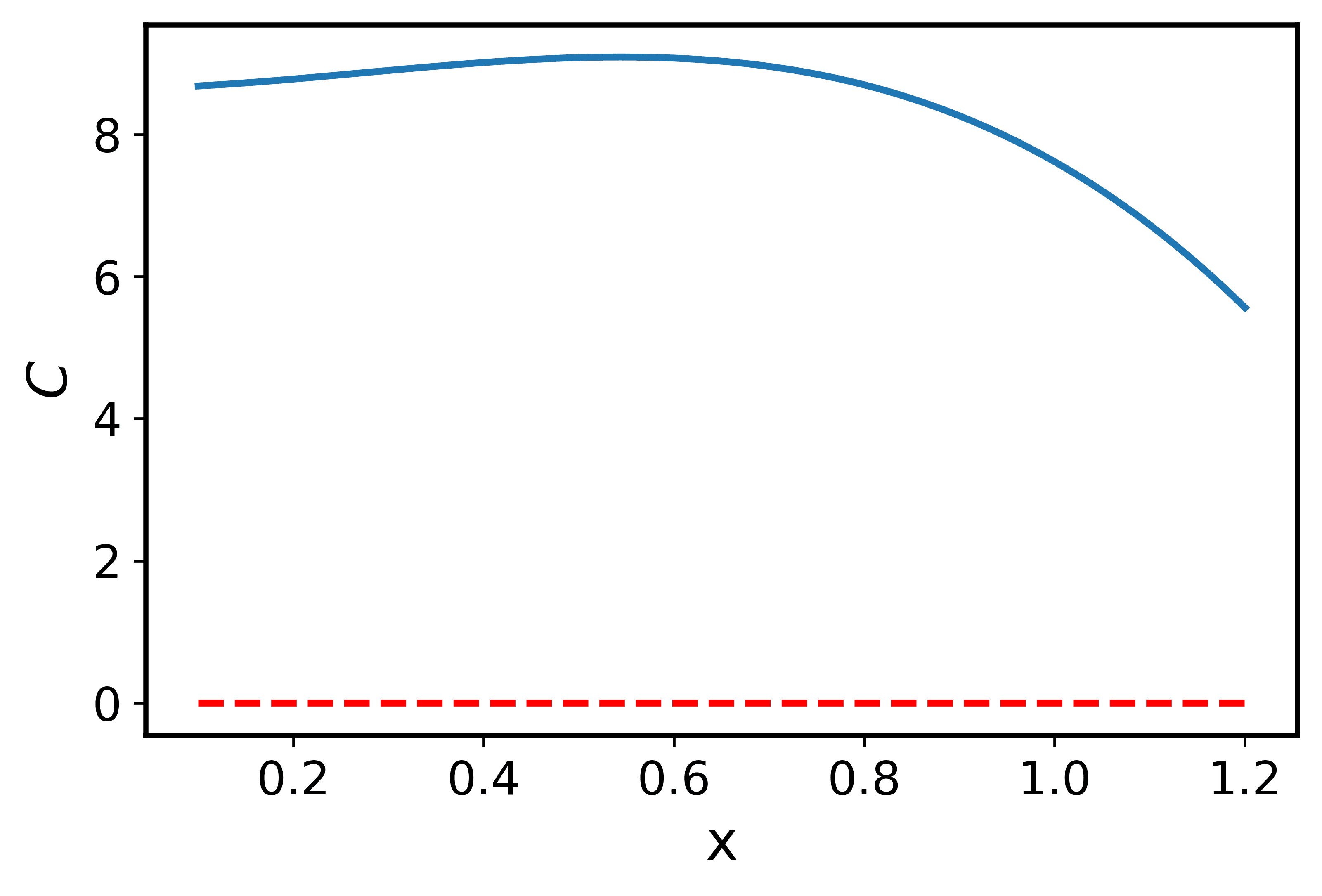}
    \end{subfigure}
    \caption{$B$ (top) and $C$ (bottom) 
for $S=5/2$. The left figures have $r_0=0$ and the right ones $r_0=5$. 
All of them have $a_1=0$.}
\label{cas_BL}
\end{figure}

In Fig.~\ref{cas_BL}, we show the parameters $B$ and $C$ for two 
situations. The plots in the left correspond to the continuous region 
(yellow) of Fig.~\ref{fig.explor}, that is with $r_0=0$ and $a_1=0$. And the 
plots in the right stand for the discontinuous region (black) of 
Fig.~\ref{fig.explor}, that is with $r_0=5$ and $a_1=0$. In the first 
case ($r_0=0$) the critical point is at 
 $x=0.63$, and in the second one ($r_0=0.5$) at $x=0.666$, according to 
Landau theory. For those critical values,  $C$ is 
positive for both situations. However, in the first scenario, $B$ is negative 
because it changes sign before reaching a value of $0.6$, but, in the second 
one, $B$ is positive because the change of sign happens after a value of $0.7$. 

\begin{figure}[tb]
    \centering
    \begin{subfigure}[b]{0.4\textwidth}
        \centering
        \includegraphics[width=\textwidth]{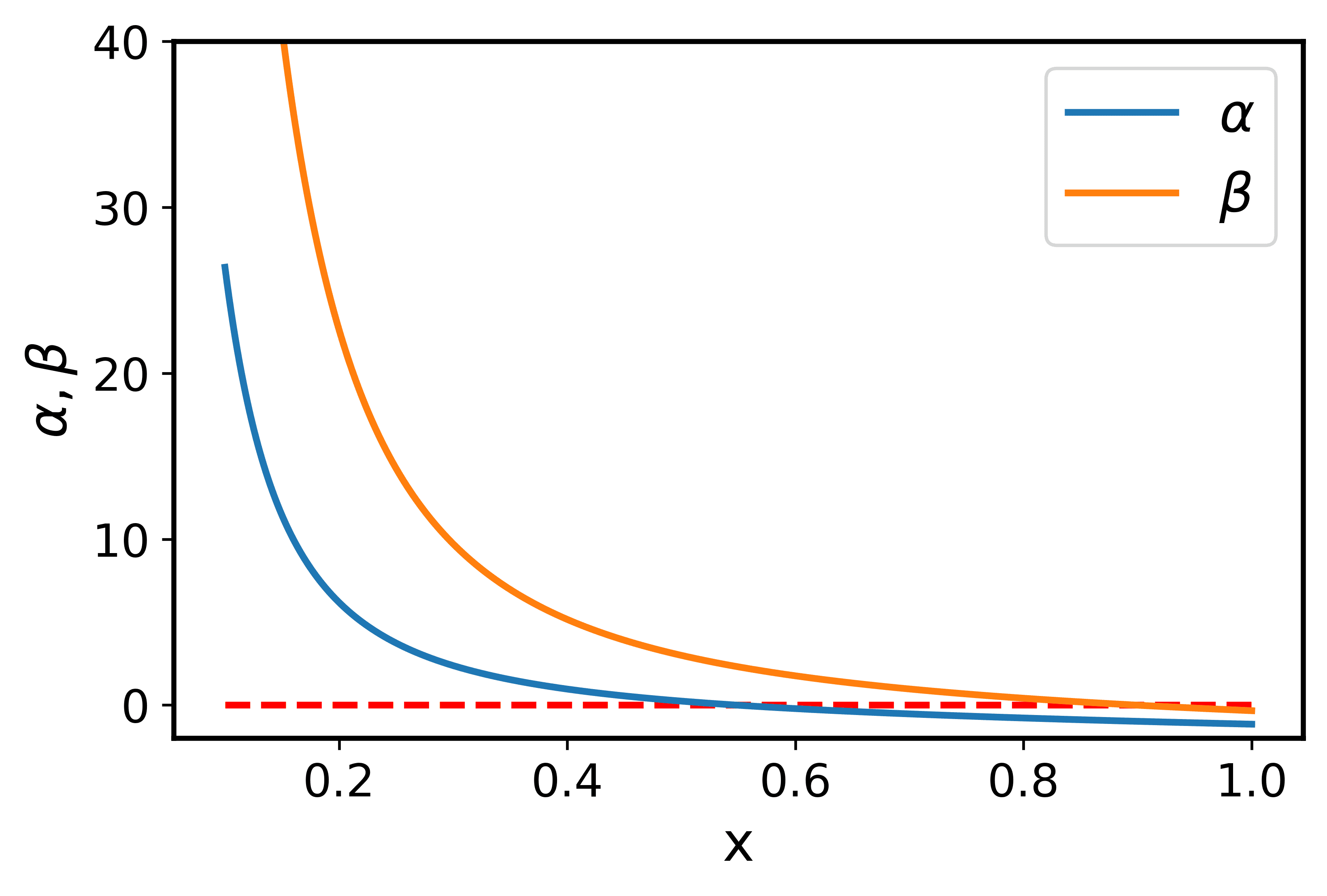}
    \end{subfigure}
    \begin{subfigure}[b]{0.4\textwidth}
        \centering
        \includegraphics[width=\textwidth]{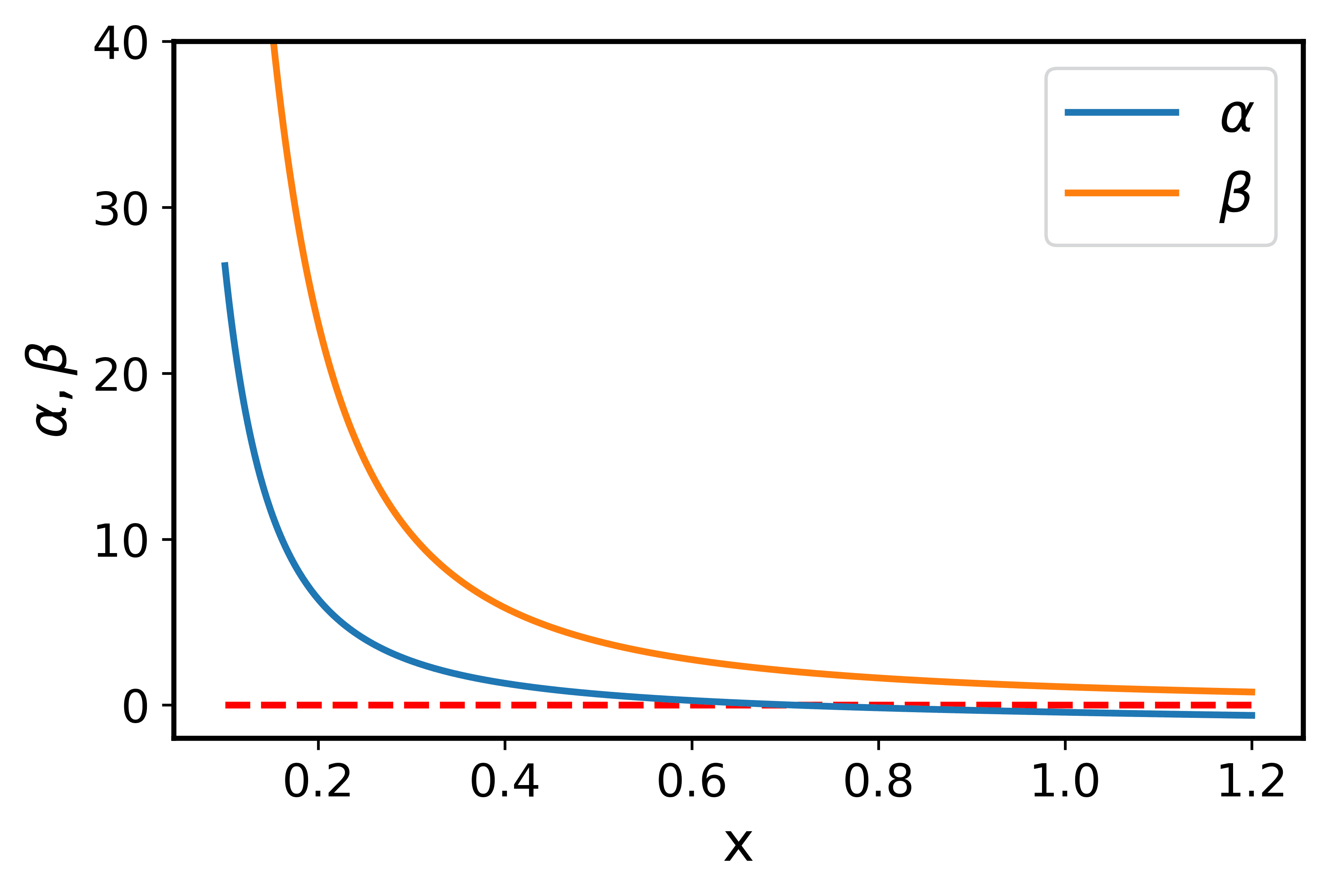}
    \end{subfigure}
    \caption{Comparison of $\alpha$ and $\beta$ in terms of $x$ 
for $r_0=0$ (top) and $r_0=5$ (bottom). Both cases stand for $S=5/2$ and 
$a_1=0$.}
\label{cas_albe}
\end{figure}

So, in the case $r_0=0$, at the transition point, $\beta$ is positive because 
both $L$ and $C$ are positive, but, $\alpha$ is negative because $B$ has changed 
sign. This can be seen in Fig.~\ref{cas_albe}. Therefore, according to the 
conditions stated above, the discontinuous transition is not possible. 
Landau theory for this case concludes that the change of sign of $B$ 
allows a continuous transition to happen at spin values larger than $1/2$. The 
expected overall negative sign in the cubic term $P^3$ is what lead to believe 
that the transition would be always discontinuous. But here, we are 
seeing that $B$ can change sign, and make the overall sign of the cubic term 
positive, eliminating the inflection in the energy, and hence, giving rise to a 
continuous transition. In 
the second scenario ($r_0=5$), at the transition point, everything is positive: 
$B$, $C$, $L$, $\alpha$, and $\beta$ (see Fig.~\ref{cas_albe}). Therefore, as $\beta$ is larger than $\alpha$ a discontinuous transition emerges, 
in agreement with perturbation theory. Essentially, this is the expected case in which the overall sign of the cubic term is negative, causing an inflection in the energy and making the transition discontinuous.

\section{Summary and Conclusions}

We have studied the itinerant ferromagnetism phase transition of Fermi 
gases as a function of the $s$-wave effective range and  $p$-wave scattering 
length within perturbation theory to third order in the gas 
parameter~\cite{Pera2}. The results presented stand for gases with $S=1/2$, 
$3/2$, $5/2$, $7/2$ 
and $9/2$. We recall that electrons have $S=1/2$, Ytterbium has $S=5/2$, and 
Strontium has $9/2$. 

The nature of this phase transition has been discussed for a long time.
First, the Stoner model~\cite{stoner} predicted a continuous transition for 
$S=1/2$, and a 
discontinuous one for the rest. Later, a second-order model predicted that all 
the transitions are discontinuous, independently of the spin value~\cite{Pera}. 
At third 
order, the description changes again: any Fermi gas, independently of its spin, 
can have any kind of transition. It 
does not matter the spin value, we can either have a continuous transition, a 
discontinuous one, or even no transition at all. Perhaps, the more striking 
result is that there can be continuous transitions for $S>1/2$~\cite{Pera2}. Up 
to now, it 
was well established that for $S>1/2$ the transition would be  discontinuous, 
and the main discussion happened around the $S=1/2$ case.

Our results show that for $S=1/2$, the dominant transition is the 
continuous one, which happens 
mostly for positive values of $r_0$. However, there exists a small region for negative 
values of $r_0$ where the discontinuous transitions emerge. On the contrary, for 
$S>1/2$, this situation is reversed. For positive values of $r_0$, we encounter 
first-order phase transitions, while for negative and small values of $r_0$, Fermi gases suffer a second-order phase transition.
In order to understand this variability, we have used Landau theory. For 
$S>1/2$, we show why there can be a continuous transition for low values of 
$r_0$, and also explain the discontinuous regime. For $S=1/2$, the 
continuous behavior has been explained, however, Landau theory has failed to 
account for the discontinuous solutions.

Incorporating the two additional scattering parameters has enriched the itinerant ferromagnetism scenario. Should someone delve deeper into perturbation theory and provide additional detail and information regarding the potential, we anticipate not a decrease of this rich variability, but rather a modification of the domains where different phase transitions emerge.

\section{Acknowledgments}
We acknowledge financial support from Ministerio de Ciencia e Innovaci\'on
MCIN/AEI/10.13039/501100011033
(Spain) under Grant No. PID2020-113565GB-C21 and
from AGAUR-Generalitat de Catalunya Grant No. 2021-SGR-01411.

\appendix
\clearpage
\begin{widetext}

\section{Parameters of the Landau expansion}

\label{appendix:b}

In this section we show the expression for all the parameters of the Landau 
expansion in terms of the gas degeneracy ($\nu=2s+1$) and the gas parameter 
($x=k_Fa_0$). We first show the Landau expansion up to fourth order in $P$ 
including the logarithmic term that comes from the second order term in 
perturbation theory,
\begin{equation}
\begin{aligned}    
f(x,P)-f_0(x)=-\frac{A}{2}(\overline{x}-x_0)P^2-\frac{B(x)}{3}{|P|}^3+\frac{C(x)
} { 4 } P^4+\frac {
L(x)}{4}P^4\ln{|P|} \ .
\end{aligned}
\end{equation}

The parameters that are invariant in any order are
\begin{equation}
\begin{aligned}
    f(x,P)=\frac{5E}{3N\epsilon_F}\quad,\quad 
A=\frac{20}{9\pi}(\nu-1)\quad\text{, and}\quad 
x_0=\pi/2 \ .
\end{aligned}
\end{equation}

We will start with the constant term with respect to $P$,
\begin{equation}
f_0(x)=1+\frac{10}{9\pi}(\nu-1)x+\frac{4}{21\pi^2}(\nu-1)(11-2\ln{2})x^2+\frac{
1 } { 6\pi}(\nu-1)x_0x^2+\frac{1}{3\pi}(\nu+1)x_1^3+(\nu-1)[ 
0.125964142+0.095637082(\nu-3)]x^3 \ ,
\end{equation}
with $x_1=k_F a_1$.

Then, the modified $x$ we have called $\overline{x}$,
\begin{equation}
    \overline{x}=x+\frac{2}{15\pi}(22-7\nu+4(\nu-1)\ln{2})x^2-\frac{\nu-4}{12}x_0x^2-\frac{\nu+4}{6}x_1^3-\frac{9\pi}{10}(0.355828883-0.139542655\nu-0.027888923\nu^2)x^3 \ .
\end{equation}

The next parameter is the one of ${|P|}^3$, we notice that this term is zero if $\nu=2$,
\begin{equation}
\begin{aligned}
   B(x)=\frac{5}{27}(\nu-1)(\nu-2)+\frac{2}{81\pi^2}
(\nu-1)(\nu-2)(44+\nu+8(\nu-1)\ln {
2})x^2+\frac{5(\nu-1)(\nu-2)(\nu+8)}{162\pi}x_0x^2\\+\frac{(\nu-1)(\nu-2)(5\nu-40)}{81\pi}x_1^3-3(\nu-1)(\nu-2)(0.086700582+0.016099453\nu+0.008647721\nu^2)x^3 \ .
\end{aligned}
\end{equation}

Now, the term of $P^4$,
\begin{equation}
\begin{aligned}  
C(x)=\frac{20}{243}(\nu-1)(\nu^2-3\nu+3)+(\nu-1)\frac{528-336\nu-16\nu^2+29\nu^3
   -(96-192\nu+128\nu^2-32\nu^3)\ln{2}}{729\pi^2}x^2\\+\frac{20\nu^3}{243\pi^2}
(\nu-1)\ln{\frac{\nu}{6}}x^2+\frac{10}{729\pi}(\nu-1)(\nu+2)(\nu^2-3\nu+3)x_0x^2+\frac{20}{729\pi}(\nu-1)(v-2)(\nu^2-3\nu+3)x_1^3+\\
+4(\nu-1)[0.114687364-0.043172957\nu+0.006693296\nu^2+0.007873117\nu^3-0.002568852\nu^4\\-(0.005175610-0.001007688\nu)\nu^3\ln{\nu}]x^3 \ .
\end{aligned}
\end{equation}

And finally, the parameter of the logarithmic term,
\begin{equation}
   L(x)=\frac{20\nu^3}{243\pi^2}
(\nu-1)x^2-4(\nu-1)v^3(0.005175610-0.001007688\nu)x^3 \ .
\end{equation}

\end{widetext}

\bibliography{refs}

\end{document}